\documentclass[preprint,showpacs,preprintnumbers,amsmath,amssymb]{revtex4}

\usepackage{graphicx}
\usepackage{dcolumn}
\usepackage{bm}

\begin{document}


\title{All-Dielectric Rod-Type Metamaterials at Optical Frequencies}

\author{K. Vynck}
\author{D. Felbacq}
 \email{felbacq@ges.univ-montp2.fr}
\author{E. Centeno}
\author{A. I. C\u{a}buz}
\author{D. Cassagne}
\author{B. Guizal}
\affiliation{Groupe d'Etude des Semiconducteurs, UMR 5650 CNRS-UM2, CC074, Place Eug\`{e}ne Bataillon, 34095 Montpellier Cedex 05, France}

\date{\today}

\begin{abstract}
Light propagation in all-dielectric rod-type metamaterials is studied theoretically. The electric and magnetic dipole moments of the rods are derived analytically in the long-wavelength limit. The effective permittivity and permeability of a square lattice of rods are calculated by homogenizing the corresponding array of dipoles. The role of dipole resonances in the optical properties of the rod array is interpreted. This structure is found to exhibit a true left-handed behavior, confirming previous experiments [L. Peng \textit{et al.}, Phys. Rev. Lett. \textbf{98}, 157403 (2007)]. A scaling analysis shows that this effect holds at optical frequencies and can be obtained by using rods made, for example, of silicon.
\end{abstract}



\maketitle
Metamaterials (MMs) are artificial structures made of microscopic elements whose collective behavior results in unusual macroscopic optical properties \cite{Smith2004, Pendry2006, Schurig2006}. Great efforts are currently being made to scale MMs down to optical frequencies \cite{Engheta2005, Zhang2006, Soukoulis2007a, Shalaev2007}. Such an achievement would indeed make possible the development of novel, sophisticated, optical technologies in many areas, including telecommunications, life sciences and solar power.

Recent works \cite{Huang2004, Wheeler2005, Jylha2006, Peng2007, Schuller2007} have made a step forward by proposing to use high-permittivity dielectric objects instead of metallic ones to avoid the losses and saturation effects inherent to the metal in the optical range \cite{Soukoulis2007b}. This approach relies on the resonant modes that dielectric objects support \cite{VandeHulst1981}. Collections of resonators are expected to strongly modify the propagation of light at frequencies close to the resonances. Previous studies have for example noticed a correlation between the resonances of single dielectric objects and the opening of photonic band gaps (PBGs) in arrays of them \cite{Lidorikis1998, Moroz1999}. In the context of MMs, arrays of dielectric rods in p-polarized light (magnetic field parallel to the axis of the rods) have been shown to possess an effective, dispersive, magnetic permeability \cite{OBrien2002, Felbacq2005}. Interestingly, it has further been suggested that dielectric rods in s-polarized light (electric field parallel to the axis of the rods) could exhibit both electric \textit{and} magnetic dipole resonances, and thus, possibly constitute a so-called left-handed (LH) medium with simultaneously negative permittivity and permeability \cite{Peng2007, Schuller2007}. These dipole resonances have been explained in terms of strong charge displacements and displacement currents, but to our knowledge, no theory on the electric and magnetic dipole activities of dielectric rods and their role in the optical properties of rod arrays has been given up to now. This, however, is an important matter if such structures are to be used in practical MM-based applications operating at optical frequencies.

In this Letter, we present a theoretical study on the optical properties of periodic arrays of dielectric rods from the point of view of MMs. Our objective is to prove that these structures offer a similar control over light as conventional metallic MMs and that typical MM properties (e.g. left-handedness) can be observed at optical frequencies in very realistic and simple designs. We proceed as follows. First, we describe the electric and magnetic dipole activities of isolated rods by deriving explicit expressions of the corresponding dipole moments in the long wavelength limit. Second, we compute the effective material parameters of a square lattice of rods from these expressions and explain how the electric and magnetic dipole resonances of the rods cause the appearance of PBGs and LH dispersion curves. Third, we show that these effects can be reproduced at different frequencies by tuning the rod resonances and illustrate our claim by the numerical demonstration of a \textit{true} LH behavior at optical frequencies.

We start by considering an isolated, infinitely long dielectric rod of circular cross-section $C$, radius $R$ and relative permittivity $\varepsilon$, surrounded by air, in a cartesian coordinate system of unit vectors $\mathbf{u}_x$, $\mathbf{u}_y$ and $\mathbf{u}_z$, where the rod axis is along the latter. A planewave of wavevector $\mathbf{k}$ ($|\mathbf{k}|=k=2\pi/\lambda$) propagating along the $x$-direction illuminates the rod. The following theory is given for s-polarized light but similar steps could be carried out in the p-polarization.

The scattering of light by circular cylinders is described by Mie theory, which provides exact analytical solutions of Maxwell's equations \cite{VandeHulst1981}. In particular, the scattered electric field $\mathbf{E}^s$ in the far zone ($kr \gg 1$) is given by:
\begin{equation}\label{E:01}
 \mathbf{E}^{s}(\mathbf{r}) = \sqrt{\frac{2}{\pi}} \frac{e^{ikr}}{\sqrt{kr}} \: e^{-i\frac{\pi}{4}} \left( b_{0} + 2 \sum_{n=1}^{+\infty} b_{n} \cos(n\theta) \right) \mathbf{u}_z
\end{equation}
where $\theta$ is the angle with respect to the $x$-direction and $b_{n}$ the $n$th-order Mie scattering coefficient of the rod. In fact, the polarization per unit volume $\mathbf{P} = \varepsilon_0 (\varepsilon-1) \: \mathbf{E}$ induced in the dielectric rod acts as a source for the scattered field. $\mathbf{E}^s$ can then be written in an integral form using Green's theorem as \cite{Felbacq1994}:
\begin{equation}\label{E:03}
 \mathbf{E}^s(\mathbf{r}) = \frac{ik^2}{4} \int_C H_0^{(1)} (k |\mathbf{r}-\mathbf{r'}|) \: (\varepsilon-1) \mathbf{E}(\mathbf{r'}) d^2r'
\end{equation}
where $H_0^{(1)}$ is the zeroth order of the Hankel function of the first kind. The integral has been restricted to $C$ because the polarization vanishes outside the rod.

A second expression of the far-field expansion in Eq.~(\ref{E:01}) can now be derived by expanding Eq.~(\ref{E:03}) into a series of multipoles. Note that this technique differs from the familiar three-dimensional multipole expansion of the magnetic vector-potential \cite{Jackson1999} because the bidimensionality of our problem necessarily implies a strong effect of the light polarization on the scattered field. In the far zone, $H_0^{(1)}$ can be expressed by its asymptotic form \cite{Abramowitz1964}. The multipole expansion is introduced by writing $|\mathbf{r}-\mathbf{r'}| \simeq r-\mathbf{u_r} \cdot \mathbf{r'}$, where $\mathbf{r}=r\mathbf{u_r}$, yielding the approximations $\sqrt{k|\mathbf{r}-\mathbf{r'}|} \simeq \sqrt{kr}$ and $e^{ik|\mathbf{r}-\mathbf{r'}|} \simeq e^{ikr} \cdot e^{-ik \mathbf{u_r} \cdot \mathbf{r'}}$. The exponential $e^{-ik \mathbf{u_r} \cdot \mathbf{r'}}$ is then expanded in series of the source extension versus the wavelength as $e^{-ik \mathbf{u_r} \cdot \mathbf{r'}} = \sum_{n=0}^{\infty} \frac {(-ik\mathbf{u_r} \cdot \mathbf{r'})^n} {n!}$. By inserting these expressions in Eq.~(\ref{E:03}), we obtain the \textit{polarized multipole expansion} of the scattered electric field in the far zone:
\begin{equation}\label{E:05}
 \mathbf{E}^{s} (\mathbf{r}) = \sqrt{\frac{2}{\pi}} \frac{e^{ikr}}{\sqrt{kr}} \: e^{-i\frac{\pi}{4}} \sum_{n=0}^{\infty} \mathbf{f}_n(\mathbf{r})
\end{equation}
with $\mathbf{f}_n(\mathbf{r}) = \frac{ik^2}{4} \frac{(-ik)^n}{n!} \int_C (\mathbf{u_r} \cdot \mathbf{r'})^n \: (\varepsilon-1) \mathbf{E}(\mathbf{r'}) d^2r'$.

The successive terms of this expression describe 2D multipole radiation fields at large distances, the zeroth ($n=0$) and first ($n=1$) orders being assimilated to electric and magnetic dipoles, respectively. The corresponding scattering orders can be written as a function of the electric and magnetic dipole moments per unit length, defined respectively by $\mathbf{p}= \int_C \mathbf{P}(\mathbf{r'}) d^2r'$ and $\mathbf{m}= \frac{1}{2} \int_C \mathbf{r'} \times \mathbf{J}(\mathbf{r'}) d^2r'$, with $\mathbf{J} = \partial \mathbf{P} / \partial t$ the polarization current density. By equating them with their counterpart in Eq.~(\ref{E:01}), we can write $\mathbf{p}$ and $\mathbf{m}$ as a function of the Mie scattering coefficients $b_0$ and $b_1$, respectively, as:
\begin{subequations}\label{E:06}
 \begin{equation}
  \mathbf{p}/\varepsilon_0=(4b_0/ik^2) \mathbf{u}_z
 \end{equation}
 \begin{equation}
  \mathbf{m} Z_0 =(-4b_1/ik^2) \mathbf{u}_y
 \end{equation}
\end{subequations}
with $Z_0=\sqrt{\mu_0/\varepsilon_0}$ the free space impedance. These two expressions describe analytically the dipole activities of isolated rods in the long-wavelength limit in terms of their scattering matrix.

\begin{figure}
\includegraphics{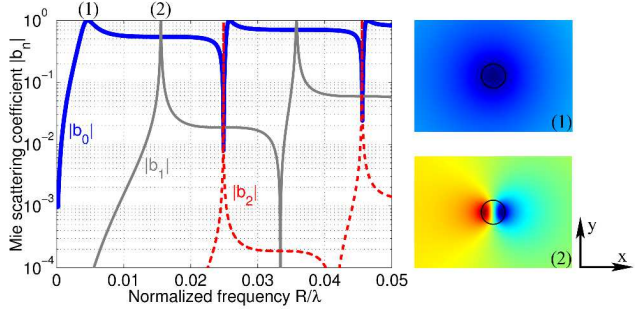}
\caption{\label{fig:1} (Color online) Complex modulus of $b_0$ (thick blue solid line), $b_1$ (thin gray solid line) and $b_2$ (thin red dashed line) for rods with permittivity $\varepsilon=600$, in logarithmic scale. The insets show the amplitude of the scattered electric field at the first maxima of $|b_0|$ and $|b_1|$.}
\end{figure}

The Mie scattering coefficients $b_0$, $b_1$ and $b_2$ are plotted in Fig.~\ref{fig:1} for rods with permittivity $\varepsilon=600$ \cite{Peng2007}. On the whole, $|b_0|$ and $|b_1|$ are found to remain larger by a few orders of magnitude than the higher-order coefficients, suggesting that the electric and magnetic dipole activities of the rods contribute to the most part of the optical properties of rod arrays. Higher-order multipoles, which are expected to induce spectrally narrow optical features near their resonance frequencies, can safely be ignored.

We now consider an array of dielectric rods. In the long-wavelength limit, the corresponding array of dipoles can be described as an effective medium with permittivity and permeability dyadics $\bar{\bar{\varepsilon}}$ and $\bar{\bar{\mu}}$, respectively. These macroscopic parameters are calculated from the microscopic polarizabilities of the rods through a process of homogenization, which takes the density of resonators and their mutual interaction into account. In the case of s-polarized light propagating along the $x$-direction, only the $\varepsilon_{zz}$ and   $\mu_{yy}$ components are required to define the refractive index of the effective medium $n_{\text{eff}}=\sqrt{\varepsilon_{zz} \mu_{yy}}$. Considering that the incident electric field amplitude has been normalized to unity and using the relations $|\mathbf{H}^i|=|\mathbf{E}^i|/Z_0$ and Eqs.~(\ref{E:06}), the electric and magnetic polarizabilities per unit length of the rods can be written as $\alpha^e_{zz}=p_z/\varepsilon_0 E_z^i={4b_0}/{ik^2}$ and $\alpha^m_{yy}=m_y/H_y^i={4b_1}/{ik^2}$, respectively.

\begin{figure}
\includegraphics{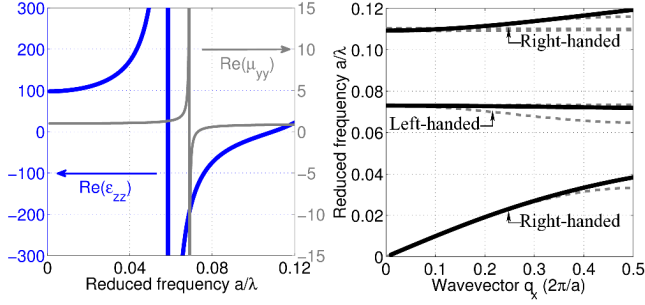}
\caption{\label{fig:2} (Color online) (Left) Real parts of the effective permittivity $\varepsilon_{zz}$ (thick blue line) and permeability $\mu_{yy}$ (thin gray line) of the rod array ($R=0.68a/3$, $\varepsilon=600$). (Right) Dispersion curves of the structure calculated from the effective material parameters (black solid lines) and by the PWE method (gray dashed lines).}
\end{figure}

The effective material parameters of a square lattice of rods with permittivity $\varepsilon=600$ and radius $R=0.68a/3$, where $a$ is the lattice periodicity, are calculated using the nonlocal homogenization model proposed by Silveirinha \cite{Silveirinha2006} in the approximation of wavevectors close to the $\Gamma$-point. For the sake of comparison, this structure is similar to the one studied in Ref.~\cite{Peng2007}. As shown in Fig.~\ref{fig:2}~(left), the first electric dipole resonance of the rods [first inset of Fig.~\ref{fig:1}] induces a strong resonance of the permittivity, which results in the opening of a wide frequency range of negative permittivity. At $a/\lambda \simeq 0.07$, the first magnetic dipole resonance of the rods [second inset of Fig.~\ref{fig:1}] induces a sharp resonance of the permeability, which eventually takes negative values within the negative permittivity range, thereby inferring a LH behavior in accordance with Peng \textit{et al.} \cite{Peng2007}. In Fig.~\ref{fig:2}~(right), we compare the dispersion curves of the effective medium calculated using the relation $q_x=n_{\text{eff}} \: \omega/c$ and those of the rod array, calculated by the planewave expansion (PWE) method using a freely available software package \cite{Johnson2001}. Apart from symmetry degeneracies and higher-order resonances, which have not been considered in our theory, the main optical features of the rod array are very well reproduced. We can then confidently say that this structure truly behaves as a medium with negative permittivity in the two lower-frequency PBGs, and as a LH medium in the narrow frequency range at $a/\lambda \simeq 0.07$.

At this point, we have shown that the optical properties of periodic arrays of rods result from the collective response of the resonant rods. These structures rely on similar principles as conventional metallic MMs and thus, may be used as such to control the propagation of light.

\begin{figure}
\includegraphics{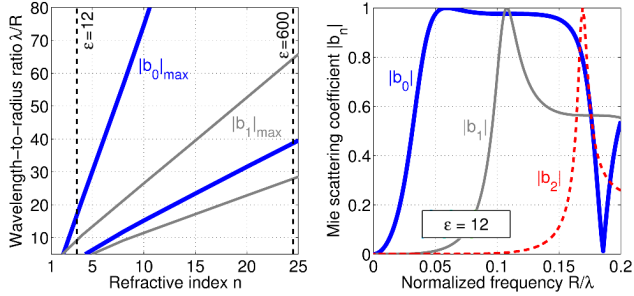}
\caption{\label{fig:3} (Color online) (Left) Wavelength positions (in units of $\lambda/R$) of the two lower-frequency maxima of $|b_0|$ (thick blue lines) and $|b_1|$ (thin gray lines) as a function of the rod refractive index $n$. (Right) Complex modulus of $b_0$ (thick blue solid line), $b_1$ (thin gray solid line) and $b_2$ (thin red dashed line) for rods with permittivity $\varepsilon=12$.}
\end{figure}

Due to the growing interest in developing MMs for the optical range, it is now important to investigate the scaling properties of all-dielectric rod-type structures. Previous studies have limited their work to high-permittivity rods to place their resonances in the homogeneous regime ($\lambda \gg a, R$) and prevent them from exhibiting a strong spatial dispersion \cite{Huang2004, Peng2007, Schuller2007}. Our theory provides additional information on this matter. As shown above, the electric and magnetic dipole activities of dielectric rods are intrinsically related to their Mie scattering coefficients. The scaling properties of rod-type structures may therefore be understood by studying the variation of these coefficients with the rod permittivity $\varepsilon$. Decreasing $\varepsilon$ reduces the optical size of the rods and thus, the wavelength $\lambda$ at which they resonate. This is evidenced in Fig.~\ref{fig:3}~(left), where the wavelength positions of the electric and magnetic dipole resonances (indicated by the maxima of $|b_0|$ and $|b_1|$) are shown to depend almost linearly on the rod refractive index $n=\sqrt{\varepsilon}$. In particular, the magnetic dipole resonance observed at $\lambda/R \simeq 63$ ($a/\lambda \simeq 0.07$) in rods with permittivity $\varepsilon=600$ is shifted to $\lambda/R \simeq 8.8$ ($a/\lambda \simeq 0.5$) when $\varepsilon=12$. As shown in Fig.~\ref{fig:3}~(right), this permittivity is sufficiently high for $b_0$ and $b_1$ to remain the most significant coefficients up to the resonance frequency of $b_2$ at $R/\lambda \simeq 0.17$. The LH behavior is therefore expected to hold.

\begin{figure}
\includegraphics{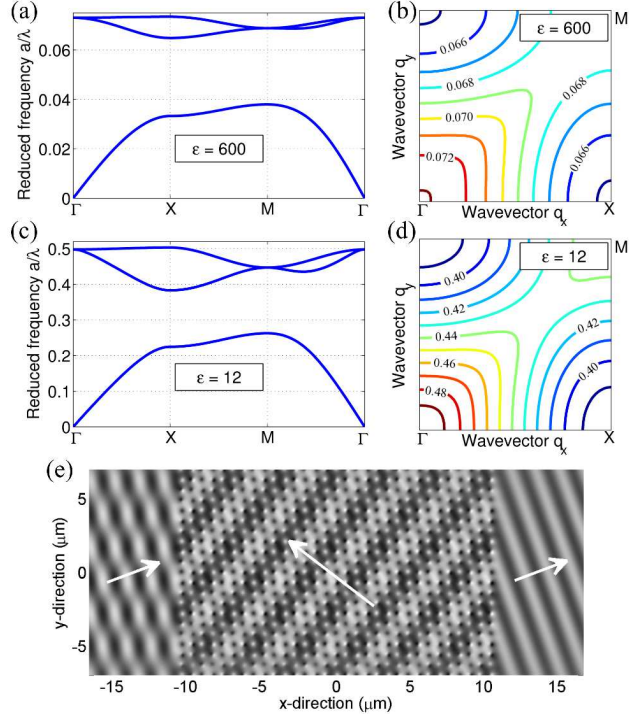}
\caption{\label{fig:4} (Color online) Photonic band structures (a, c) and second-band IFCs (b, d) of the higher and lower-$\varepsilon$ rod arrays (top: $\varepsilon=600$, down: $\varepsilon=12$). The IFCs are labeled by their reduced frequency $a/\lambda$. (e) Steady-state amplitude of the electric field of light at $\lambda=1.55$ $\mu$m incident at $20^\circ$ on the lower-$\varepsilon$ structure ($R=158$ nm, $a=698$ nm).}
\end{figure}

This is verified by comparing the photonic band structures and second-band iso-frequency curves (IFCs) of the higher and lower-$\varepsilon$ structures calculated by the PWE method. As observed in Figs.~\ref{fig:4}(a)-(d), their optical features are readily similar. First, the LH dispersion curves lying at $a/\lambda \simeq 0.07$ are pushed up to $a/\lambda \simeq 0.5$, as expected. The splitting between the two higher-frequency bands is larger in the latter structure because the decrease of the rod refractive index broadens the rod resonances [Fig.~\ref{fig:3}~(right)] and increases the inter-rod interaction \cite{Lidorikis1998}. Second, the IFCs both exhibit a strong spatial dispersion even in the vicinity of the $\Gamma$-point. This observation supports previous studies \cite{Belov2003, Cabuz2007}, arguing that large wavelength-to-period ratios do not necessarily result in isotropic responses. The spatial dispersion here is naturally inferred by the magnetic dipole activity of the structure \cite{Landau1984, Agranovich2004}. The lattice symmetry therefore plays an important role in the spatial response of the rod array in the LH frequency range.

To evidence the existence of a LH behavior at optical frequencies, we finally simulate the propagation of light in the lower-$\varepsilon$ structure with the 2D finite-difference time-domain method using a freely available software package \cite{Farjadpour2006}. Light is incident at an angle of $20^\circ$ and the structure is tuned to near-infrared wavelengths ($\lambda=1.55$ $\mu$m) by using rods of radius $R=158$ nm and a lattice of periodicity $a=698$ nm. As shown in Fig.~\ref{fig:4}(e), the phase of the propagating field in the rod array is opposite to that of the field in free space, which indicates, as expected, a LH behavior.

Taking all the above into consideration, we come to the conclusion that the LH behavior initially observed by Peng \textit{et al.} \cite{Peng2007} is \textit{not} specific to large wavelength-to-period ratios, for this effect is primarily a matter of coupled resonances.

It is particularly important to note that the physics involved here remains the same even though, in lower-$\varepsilon$ structures, $\lambda$ becomes comparable to $a$. In this regard, it would be interesting in future studies to investigate to what extent these structures can be considered as homogeneous. Concurrently, our theoretical results could be used to clarify the origin of the many LH behaviors that have been reported in rod-type photonic crystals in the past few years (see e.g. Ref.~\cite{Foteinopoulou2003}). Experiments could also be initiated at once, as techniques to fabricate silicon-based ($\varepsilon \simeq 12$) rod-type structures already exist \cite{Xu2001}. Silicon holds a pre-eminent position in Photonics and is therefore a very interesting candidate material for all-dielectric MMs operating at optical frequencies. Rod-type structures are finally likely to integrate well on photonic platforms, considering that high coupling efficiencies can be obtained even at large angles of incidence owing to the rod resonances \cite{Botten2006}.

To conclude, we have presented a theoretical study on the optical properties of all-dielectric rod-type structures, using an approach based on the electric and magnetic dipole activities of dielectric rods. This work constitutes a first proof that dielectric rods can be used to design true MMs at optical frequencies. Due to the large experimental knowledge in fabricating nanoscale dielectric structures, exciting applications, such as all-dielectric invisibility cloaks, may be realized in the optical domain in the near future.

\begin{acknowledgments}
The support of the european NoE project Nr 511616 PhOREMOST and the french ANR project Nr 06-0030 POEM-PNANO are gratefully acknowledged.
\end{acknowledgments}

\end{document}